\documentclass[a4paper]{article}
\usepackage[a4paper,top=3cm,bottom=2cm,left=3cm,right=3cm,marginparwidth=1.75cm]{geometry}
\usepackage{amsfonts}
\usepackage{bm}
\usepackage{extarrows}
\usepackage{amssymb}
\usepackage{color}
\usepackage[all]{xy}
\usepackage{graphicx}
\usepackage{braket}
\usepackage{amsmath}
\usepackage{appendix}
\usepackage{graphicx}
\usepackage[colorinlistoftodos]{todonotes}
\usepackage{amsthm}
\usepackage{mathrsfs}
\usepackage{amssymb}
\usepackage{accents}
\usepackage{extarrows}
\usepackage{makecell}
\usepackage{authblk}
\usepackage{url}
\usepackage{graphicx}
\usepackage{hyperref}
\usepackage{cite}
\usepackage{eufrak}
\hypersetup{colorlinks,
            citecolor=black,
            linkcolor=black,
            urlcolor=green,
            pdftex}

\def\be{\begin{equation}}
\def\ee{\end{equation}}
\def\ba{\begin{eqnarray}}
\def\ea{\end{eqnarray}}

\usepackage[english]{babel}
\usepackage[utf8x]{inputenc}
\usepackage[T1]{fontenc}

\usepackage[normalem]{ulem}

\title{Energy conditions in new model of loop quantum cosmology}
\author[1,2]{Gaoping Long \footnote{201731140005@mail.bnu.edu.cn}}
\author[1]{Yunlong Liu}
\author[1]{Xiangdong Zhang \thanks{Corresponding author. scxdzhang@scut.edu.cn}}

\affil[1]{Department of Physics, South China University of Technology, Guangzhou 510641, China}
\affil[2]{Department of Physics, Beijing Normal University, Beijing 100875, China}

\date{}

\begin{document}

\maketitle

\begin{abstract}

Recently, a de-Sitter epoch has been found in the new model of loop quantum cosmology which is governed by the scalar constraint with both of Euclidean and Lorentz terms. The singularity free bounce in the new LQC model and the emergent cosmology constant strongly suggest that the effective stress energy tensor induced by quantum corrections must violate the standard energy conditions. In this paper, we do an explicit calculation to analyze the behaviours of specific representative energy conditions, i.e., average null, strong and dominant energy conditions. It turns out that the averaging null energy condition is violated while the dominant energy condition is violated only at a period around the bounce point. More specifically, the strong energy condition is violated not only at a period around the bounce point, but also the whole period from the bounce point to the classical phase corresponding to the de Sitter period. Our results is expected to shed some lights on construction of the wormhole and time machine which usually need exotic matters violate energy conditions.
\end{abstract}

\section{Introduction}
 As the origination of our universe predicted by the standard cosmology model based on classical general relativity (GR), the singular big bang point strongly indicate that a quantum gravity theory is necessary to describe the spacetime of early universe at small scale.  Background independent and nonpertubative loop quantum gravity (LQG) is constructed as one of the candidates of quantum gravity theory, and it predicts that the spatial geometry is discrete at Planck scale \cite{Ashtekar2012Background}\cite{RovelliBook2}\cite{Han2005FUNDAMENTAL}\cite{thiemann2007modern}. This discrete geometry feature is inherited in the loop quantum cosmology (LQC), which is given by studying the loop quantum states of homogeneous and isotropic geometry \cite{Bojowald:2006da}\cite{Ashtekar:2011ni}\cite{Agullo:2016tjh}. The LQC model reproduces the standard Wheeler-DeWitt theory for cosmology at large scale, while leads to notable deviations at small scale because it involves the discreteness nature of geometry. More explicitly, by applying techniques developed in the full theory of LQG, the classical Hamiltonian constraint in LQC is reformulated in terms of a new set of variables. Then the loop quantization procedure is performed to give the quantum Hamiltonian constraint, which takes the formulation as a quantum difference equation so that it evolves non-singularly through the big bang point.

The LQC model takes various formulations by following different regularization process of Hamiltonian constraint. In standard LQC, the quantization of the Hamiltonian constraint follows a standard method. Note that the Hamiltonian constraint in full theory of connection formulation of GR contains the so-called Euclidean term and Lorenzian term, which proportion to each other in the spatial-flat cosmology model classically. Usually, based on this classical equivalence, the Hamiltonian constraint in standard LQC is simplified before quantization so that it only contains the quantized Euclidean term. With this simplifying, the dynamics governed by the quantum Hamiltonian constraint predicts the big bang is replaced by a big bounce, which divides the evolution of the universe as two symmetry periods. However, it has been shown that the Euclidean term and Lorenzian term Hamiltonian constraint are not proportion to each other in the full theory of LQG\cite{2018Emergent}\cite{2019Emergent}\cite{Han:2019vpw}. By dealing these two terms separately through Thiemann regularization, a new model of LQC is obtained\cite{2018Emergent}\cite{2019Emergent}\cite{Yang:2009fp}. In this model, the quantization of Hamiltonian constraint gives a new fourth order difference equation rather than the second order one in standard LQC. A remarkable prediction of the dynamics given by new effective Hamiltonian constraint is that the big bounce divides the evolution of the universe as two non-symmetric periods, with one of periods being consistent with the evolution described by standard cosmology at large scale, while another one of periods is described by a standard cosmology with the presence of a large cosmological constant at large scale.

 The amazing result of the new LQC model indicates that the quantum geometry corrections are not only introduced near the region where the singularity appears, but also emerge as an additional cosmology constant in dynamics equation at large scale.
 According to the singularity theorem proved by Hawking and Penrose, the singularity is inevitable in GR if the Universe satisfies some energy conditions. Therefore the bounce in the new LQC model strongly suggests the violation of energy conditions with respect to some effective matter field. Besides, the emergent positive cosmology constant in the de Sitter epoch also contributes to the behaviour of energy conditions. It is then natural to ask: which specific energy condition is violated in the new model of LQC? where do they occur?  In current paper, we choose some representative energy conditions such as average null, strong and dominant energy conditions, to do an explicit calculation to analyze the behaviours of the energy conditions with respect to the effective stress-energy tensor given by the effective dynamics equation in the new model of LQC.  In fact, such task has been discussed in standard LQC as well as in the quantum corrected Schwarzschild-Kruskal space-time \cite{2009Averaged}\cite{Xiong:2006ey}\cite{Ashtekar:2018cay}. It has been shown that, with respect to the effective stress-energy tensor, the average null energy condition is violated in the massless scalar field coupled model, while the strong energy condition is violated only at small scale in standard LQC. With the new model of LQC introducing the cosmology constant at large scale as a new feature comparing with the standard one, we hope this feature can also be reflected in the behaviours of energy condition with respect to the effective stress-energy tensor. This is the main goal of this paper.

The following sections of this paper is organized as follows. In Sec. 2, we review some basic building blocks of the new LQC model and fix the notations we used in this paper. Then we discuss the violation of energy conditions in Sec. 3, we dealing the average null, strong and dominant energy conditions in different subsections. Conclusions and outlook are given in the last section.

\section{Effective dynamics of the new model of LQC}
Classically, the flat universe that we concern in this paper is described by the Friedman-Lemaitre-Robertson-Walker(FLRW) metric
\begin{equation}
ds^2=-dt^2+a^2(dx^2+dy^2+dy^2),
\end{equation}
with $a$ being the scale factor of the universe which only depends
on $t$ based on the homogeneity assumption of the universe. The dynamics of this system is governed by the Hamiltonian given by\cite{Bojowald:2006da}\cite{Ashtekar:2011ni}\cite{Agullo:2016tjh}
\begin{equation}
H_{\text{cl}} = -\frac{3}{8\pi G\gamma^2}\sqrt{p}c^2+H_{\text{matter}}(p_\phi,\phi),
\end{equation}
where we use the conjugate variables $(c,p)$ defined by  $c:=\gamma\dot{a}$, $p:=a^2$, where $\gamma$ represents the Barbero-Immirzi parameter, and $H_{\text{matter}}(p_\phi,\phi)$ is the Hamiltonian of matter field with $\phi$ an $p_\phi$ representing the matter field and its conjugate momentum respectively. The pair $(c,p)$ are usually used to coordinatize the phase space of LQC with Poisson bracket $\{c,p\}=\frac{8\pi G\gamma}{3}$. The dynamics equation of classical cosmology is described by the following Friedmann and Raychaudhuri equations as
\ba
H^2&=&\frac{8\pi G}{3}\rho,\\
\frac{\ddot{a}}{a}&=&\dot{H}+H^2 =-\frac{4\pi G}{3}(\rho_\phi+3P_\phi).
\ea
where $H:=\frac{\dot{a}}{a}$ is the Hubble parameter, $\rho_\phi$ and $P_\phi$ are the energy density and pressure of matter defined as follows
\ba\label{rhoP}
\rho_\phi &=&a^{-3}H_{\text{matter}},\\
P_\phi&=&-\frac{1}{3} a^{-2}\frac{\partial H_{\text{matter}}}{\partial a}
\ea
with the stress-energy tensor of the matter field takes the form
\ba
T_{\mu\nu}=\rho_\phi U_\mu U_\nu+P_\phi(g_{\mu\nu}+U_\mu U_\nu) =\rho_\phi(dt)_\mu(dt)_\nu +a^2P_\phi[(dx)_\mu(dx)_\nu+(dy)_\mu(dy)_\nu+(dz)_\mu(dz)_\nu],
\ea
where $U_\mu=(1,0,0,0)$ is the natural comoving observer of matter
field in the universe.

The new LQC model involving Lorentzian term with Thiemann regularization gives a different effective Hamiltonian constraint. In order to simplify the specific equations, we use the new canonical variables $(b,V)$ which is given by
  \begin{equation}\label{bV}
b:=c\bar{\mu},\quad V:=p^{3/2},\quad \{b,V\}=\frac{2\alpha}{\hbar},
  \end{equation}
  with $\alpha=2\pi G\hbar\gamma\sqrt{\Delta}$ and $\bar{\mu}$ being the regularization which is used in the so-call $\bar{\mu}$-scheme and defined by
  \begin{equation}
  \bar{\mu}:=\frac{\sqrt{\Delta}}{\sqrt{|p|}},\quad \Delta:=2\pi\sqrt{3}\gamma G\hbar\approx2.61\ell_{\text{Pl}},
  \end{equation}
  where $\ell_{\text{Pl}}$ is the Planck length and $\Delta$ is the smallest non-vanishing area eigenvalue from the full theory. In terms of phase space conjugated variables $(b,V)$ and $(\phi,p_\phi)$ with $\{\phi,p_\phi\}=1$, the equations of motion of the new model with Lorentzian term given by Thiemann's regularization (TR) can be derived by Hamilton's equation of the effective constraint\cite{2018Emergent}\cite{2019Emergent}\cite{Yang:2009fp}, which reads
\begin{equation}
C^{\text{TR}}_{\text{eff}}=\frac{p_\phi^2}{2V}-\frac{3}{8\pi G\Delta\gamma^2}V\sin^2(b)[1-(1+\gamma^2)\sin^2(b)]
\end{equation}
with the $H_{\text{matter}}:=\frac{p_\phi^2}{2V}$.
Then,  we have the equations of motion
\begin{equation}\label{em1}
\dot{V}=\{V,C^{\text{TR}}_{\text{eff}}\}=\frac{3}{\gamma\sqrt{\Delta}}V\sin(b)\cos(b)[1-2(1+\gamma^2)\sin^2(b)],
\end{equation}
\begin{equation}\label{em2}
\dot{b}=\{b,C^{\text{TR}}_{\text{eff}}\}=-2\pi G\gamma\sqrt{\Delta}\frac{p_\phi^2}{V^2}-\frac{3}{2\gamma\sqrt{\Delta}}\sin^2(b)[1-(1+\gamma^2)\sin^2(b)],
\end{equation}
and
\begin{equation}\label{em3}
\dot{\phi}=\{\phi,C^{\text{TR}}_{\text{eff}}\}=\frac{p_\phi}{V},\quad \dot{p_\phi}=\{p_\phi,C^{\text{TR}}_{\text{eff}}\}=0.
\end{equation}
By using the constraint equation $C^{\text{TR}}_{\text{eff}}\approx0$, we have
\begin{equation}\label{sin2b}
\sin^2b=\frac{1\pm\sqrt{1-\rho_\phi/\rho_{\text{c}}^{\text{TR}}}}{2(1+\gamma^2)}
\end{equation}
with $\rho_\phi=\frac{p_\phi^2}{2V^2}$ and $\rho_{\text{c}}^{\text{TR}}=\frac{3}{32\pi G\Delta\gamma^2(1+\gamma^2)}$ being the critical energy density of the new model. Now, we can give the modified Friedmann equation and Raychaudhuri equation as
\begin{equation}
H^2=(\frac{\dot{V}}{3V})^2=\frac{1}{\gamma^2\Delta}f(\rho_\phi) (1-f(\rho_\phi)) [1-\rho_\phi/\rho_{\text{c}}^{\text{TR}}],
\end{equation}
and
\begin{eqnarray}
\frac{\ddot{a}}{a}=\dot{H}+H^2
&=&\frac{1}{\gamma^2\Delta}\frac{\delta f(\rho_\phi) }{\delta \rho_\phi}(\frac{3}{2}(\rho_\phi+P_\phi)) (2f(\rho_\phi)-1) [1-\rho_\phi/\rho_{\text{c}}^{\text{TR}}]\\\nonumber
&&+\frac{1}{\gamma^2\Delta}f(\rho_\phi) (1-f(\rho_\phi)) [1+\frac{3}{2}P_\phi/\rho_{\text{c}}^{\text{TR}}+\frac{1}{2}\rho_\phi/\rho_{\text{c}}^{\text{TR}}]
\end{eqnarray}
with $f(\rho_\phi):=\sin^2b=\frac{1\pm\sqrt{1-\rho_\phi/\rho_{\text{c}}^{\text{TR}}}}{2(1+\gamma^2)}$ and $\frac{\delta f(\rho_\phi) }{\delta \rho_\phi}=\mp\frac{1}{4\rho_{\text{c}}^{\text{TR}}(1+\gamma^2)\sqrt{1-\rho_\phi/\rho_{\text{c}}^{\text{TR}}}},$
 where we used that $\dot{\rho}_\phi +3H(\rho_\phi+ P_\phi) =0$. This equations are regarded as the effective dynamic equations of the new model of LQC, where the quantum geometry correction and the true matter field (the massless scalar field $\phi$) can be combined as effective matter fields on GR background.
Comparing with the standard Friedmann equation and Raychaudhuri equation, we can give the effective energy density and pressure as
\begin{equation}
\rho_{\text{eff}}=\frac{3}{8\pi G\gamma^2\Delta}f(\rho_\phi) (1-f(\rho_\phi)) [1-\rho_\phi/\rho_{\text{c}}^{\text{TR}}],
\end{equation}
\begin{eqnarray}
P_{\text{eff}}=-\frac{3}{8\pi G\gamma^2\Delta}(\frac{\delta f(\rho_\phi) }{\delta \rho_\phi}(\rho_\phi+P_\phi) (2f(\rho_\phi)-1) [1-\rho_\phi/\rho_{\text{c}}^{\text{TR}}]+f(\rho_\phi) (1-f(\rho_\phi)) [1+P_\phi/\rho_{\text{c}}^{\text{TR}}]),
\end{eqnarray}
with the stress-energy tensor of the effective matter field takes the form
\begin{equation}
T^{\text{eff}}_{\mu\nu} =\rho_{\text{eff}}(dt)_\mu(dt)_\nu +a^2P_{\text{eff}}[(dx)_\mu(dx)_\nu+(dy)_\mu(dy)_\nu+(dz)_\mu(dz)_\nu].
\end{equation}
Further, let us combine the equations of motion \eqref{em2}, \eqref{em3} and Eq.\eqref{sin2b}, and note that $\rho_\phi=\frac{p_\phi^2}{2V^2}$, we can get $\sin^2b$ as a function of $\phi$ as
\begin{equation}
\sin^2(b(\phi))=\frac{1}{1+\gamma^2\cosh^2(\sqrt{12\pi G}(\phi-\phi_0))}.
\end{equation}
 Then the evolutions of the variables along the physical time $\phi$ reads
 \begin{equation}
\rho_\phi=\rho_\phi(\phi)=\frac{3}{8\pi G\Delta}[\frac{\sinh(\sqrt{12\pi G}(\phi-\phi_0))}{1+\gamma^2\cosh^2(\sqrt{12\pi G}(\phi-\phi_0))}]^2,
\end{equation}
\begin{equation}
V=V(\phi)=\sqrt{\frac{4\pi G\Delta p_\phi^2}{3}}\frac{1+\gamma^2\cosh^2(\sqrt{12\pi G}(\phi-\phi_0))}{|\sinh(\sqrt{12\pi G}(\phi-\phi_0))|},
\end{equation}
where the coordinate time $t$ and physical time $\phi$ are linked by $\frac{dt}{d\phi}=V/p_\phi$ given by
\begin{equation}
\frac{dt}{d\phi}=\text{sgn}(p_\phi)\sqrt{\frac{4\pi G\Delta}{3}}\frac{1+\gamma^2\cosh^2(\sqrt{12\pi G}(\phi-\phi_0))}{|\sinh(\sqrt{12\pi G}(\phi-\phi_0))|}.
\end{equation}
The integration of this equation is performed independently on two domains $\phi>\phi_0$ or $\phi<\phi_0$ with result being given by
\begin{equation}
t(\phi)=t_0+\frac{\gamma^2\text{sgn}(p_\phi(\phi-\phi_0))}{\sqrt{12\pi G}}[\cosh(\sqrt{12\pi G}(\phi-\phi_0))-(1+\gamma^2)\log|\coth(\sqrt{3\pi G}(\phi-\phi_0))|].
\end{equation}
From this equation, one can see that the physical time in the two domain $\phi>\phi_0$ and $\phi<\phi_0$ gives a double cover of the cosmic time $t$, and these two covers are linked by time reflection symmetry. Hence  we can focus on the domain $\phi>\phi_0$ without loss of generality. In this chart, from the point of view of a comoving observer (whose proper time is cosmic time $t$), the infinite past and  infinite future correspond to $\phi\to\phi_0^+$ and $\phi\to+\infty$ respectively. A more explicit discussion shows that, for such an observer, the far past consists of a quantum region in which the Universe is undergoing a de Sitter contracting phase dominated by an emergent cosmological constant, while the far future is given by a classically expanding phase dominated by the matter (scalar field).

The de Sitter epoch and the bounce in the new LQC model strongly suggests the violation of the energy conditions, which means, the effective stress energy
tensor $T_{\text{eff}}$ given by the effective dynamic equation of the new model of LQC must violate some energy conditions. It is then natural to ask: which specific energy condition is violated in the new model of LQC? where do they occur? Now let us discuss these issues in detail. In order to simplify sepcific expressions, let us firstly give some new inventions, which reads
\begin{equation}
\Omega_1(\phi):=\sinh(\sqrt{12\pi G}(\phi-\phi_0)),\quad \Omega_2(\phi):=\cosh(\sqrt{12\pi G}(\phi-\phi_0)),
\end{equation}
then we have
\begin{equation}\label{111}
\rho_\phi=\rho_\phi(\phi)=\frac{3}{8\pi G\Delta}[\frac{\Omega_1(\phi)}{1+\gamma^2\Omega_2^2(\phi)}]^2,
\end{equation}
\begin{equation}\label{222}
V=V(\phi)=\sqrt{\frac{4\pi G\Delta p_\phi^2}{3}}\frac{1+\gamma^2\Omega_2^2(\phi)}{|\Omega_1(\phi)|},
\end{equation}
\begin{equation}\label{333}
\frac{dt}{d\phi}=\text{sgn}(p_\phi)\sqrt{\frac{4\pi G\Delta}{3}}\frac{1+\gamma^2\Omega_2^2(\phi)}{|\Omega_1(\phi)|},
\end{equation}
\begin{equation}\label{444}
 f(\rho_\phi)=\frac{1}{1+\gamma^2\Omega_2^2(\phi)},
\end{equation}
 and
\begin{equation}\label{555}
\frac{\delta f(\rho_\phi) }{\delta \rho_\phi}=\frac{1}{4\rho_{\text{c}}^{\text{TR}}(1+\gamma^2)(1-\frac{2(1+\gamma^2)}{1+\gamma^2\Omega_2^2(\phi)})}.
\end{equation}
Here we note that the energy density $\rho_\phi$ takes value $\rho_\phi=\rho_{\text{c}}^{\text{TR}}$ when $\Omega_2^2(\phi)= \frac{1+2\gamma^2}{\gamma^2}$.
\section{The energy conditions of the new model of LQC}
Before turning to specific calculation, we should notice that $ \rho_\phi=P_\phi$ based on their definition \eqref{rhoP} with $H_{\text{matter}}:=\frac{p_\phi^2}{2V}$. Besides, the value of $\gamma$ used in numerical calculation takes the ones determined by consistency of the black hole entropy calculation, which is given by $\gamma\approx0.23753295796592$ in Ashtekar-Baez-Corichi-Krasnov (ABCK) and Domagala-Lewandowski (DL) approach \cite{Domagala:2004jt}\cite{2004Black}\cite{2010Detailed}, or $\gamma\approx0.273985635$ in Ghosh and Mitra
(GM) as well as Engle, Noui and Perez (ENP) approach \cite{Ghosh:2006ph}\cite{Kaul:1998xv}\cite{Engle:2009vc}. Then, at the critical point where $\rho_\phi=\rho_{\text{c}}^{\text{TR}}$, we have $\Omega_2^2(\phi)= \frac{1+2\gamma^2}{\gamma^2}=6.2099421$ for $\gamma=0.23753295796592$ and $\Omega_2^2(\phi)= \frac{1+2\gamma^2}{\gamma^2}=5.64982639$ for $\gamma=0.273985635$.
\subsection{The averaged null energy condition}
The averaged null energy condition for the effective model is given by
\begin{equation}
\int_{\sigma}T^{\text{eff}}_{\mu\nu}k^\mu k^\nu dl\geq0,
\end{equation}
where the integral is along arbitrary complete and achronal null geodesic $\sigma$, $k^\mu$ denotes the geodesic tangent vector, and $l$ is an affine parameter.
Following the analysis in \cite{2009Averaged}, for the homogenious and isotropic universe considered here, we choose $\sigma$ with affine parameter $l$ and tangent vector $k^\mu=(\frac{\partial}{\partial l})^\mu=\frac{1}{a}(\frac{\partial}{\partial t})^\mu+\frac{1}{a^2}(\frac{\partial}{\partial x})^\mu$ without loss of generality.
Then we can immediately give the average null energy condition as
\begin{eqnarray}\label{null}
\int_{\sigma}T^{\text{eff}}_{\mu\nu}k^\mu k^\nu dl&=&\int_{-\infty}^{+\infty}\frac{1}{a}(\rho_{\text{eff}}+P_{\text{eff}})dt\\\nonumber
&=&\int_{\phi_0}^{+\infty}\frac{-\frac{3\rho_\phi}{4\pi\gamma^2\Delta}(\frac{\delta f(\rho_\phi) }{\delta \rho_\phi} (2f(\rho_\phi)-1) [1-\rho_\phi/\rho_{\text{c}}^{\text{TR}}]+f(\rho_\phi) (1-f(\rho_\phi)) /\rho_{\text{c}}^{\text{TR}})}{\sqrt[3]{V}}\frac{dt}{d\phi}d\phi,
\end{eqnarray}
Notice that
\begin{eqnarray}
\frac{\delta f(\rho_\phi) }{\delta \rho_\phi} (2f(\rho_\phi)-1) [1-\rho_\phi/\rho_{\text{c}}^{\text{TR}}]&=& \frac{(1+\gamma^2\Omega_2^2(\phi)-2(1+\gamma^2))}{4\rho_{\text{c}}^{\text{TR}}(1+\gamma^2)} \frac{1-\gamma^2\Omega_2^2(\phi)}{(1+\gamma^2\Omega_2^2(\phi))^2}\\\nonumber
&=&(\frac{1-\gamma^2\Omega_2^2(\phi)} {4\rho_{\text{c}}^{\text{TR}}(1+\gamma^2)(1+\gamma^2\Omega_2^2(\phi))}-\frac{1-\gamma^2\Omega_2^2(\phi)}{2\rho_{\text{c}}^{\text{TR}} (1+\gamma^2\Omega_2^2(\phi))^2}),
\end{eqnarray}
and
\begin{equation}
f(\rho_\phi) (1-f(\rho_\phi)) /\rho_{\text{c}}^{\text{TR}}=\frac{\gamma^2\Omega_2^2(\phi)}{(1+\gamma^2\Omega_2^2(\phi))^2}/\rho_{\text{c}}^{\text{TR}}.
\end{equation}
Then, the integrand in \eqref{null} can be simplified as
\begin{eqnarray}
&&\frac{-\frac{3\rho_\phi}{4\pi\gamma^2\Delta}(\frac{\delta f(\rho_\phi) }{\delta \rho_\phi} (2f(\rho_\phi)-1) [1-\rho_\phi/\rho_{\text{c}}^{\text{TR}}]+f(\rho_\phi) (1-f(\rho_\phi)) /\rho_{\text{c}}^{\text{TR}})}{\sqrt[3]{V}}\frac{dt}{d\phi}\\\nonumber
&=&-\frac{3\text{sgn}(p_\phi)\rho_\phi}{4\pi\gamma^2\Delta \sqrt[3]{|p_\phi|}}(\frac{1-\gamma^2\Omega_2^2(\phi)} {4\rho_{\text{c}}^{\text{TR}}(1+\gamma^2)(1+\gamma^2\Omega_2^2(\phi))}-\frac{1-3\gamma^2\Omega_2^2(\phi)}{2\rho_{\text{c}}^{\text{TR}} (1+\gamma^2\Omega_2^2(\phi))^2})(\sqrt{\frac{4\pi G\Delta}{3}}\frac{1+\gamma^2\Omega_2^2(\phi)}{|\Omega_1(\phi)|})^{2/3}\\\nonumber
&=&-\frac{9\text{sgn}(p_\phi)}{32\pi^2G\gamma^2\rho_{\text{c}}^{\text{TR}}\Delta^2 \sqrt[3]{|p_\phi|}}(\sqrt{\frac{4\pi G\Delta}{3}})^{2/3}(\frac{1-\gamma^2\Omega_2^2(\phi)} {4(1+\gamma^2)(1+\gamma^2\Omega_2^2(\phi))}-\frac{1-3\gamma^2\Omega_2^2(\phi)}{2 (1+\gamma^2\Omega_2^2(\phi))^2})(\frac{|\Omega_1(\phi)|}{1+\gamma^2\Omega_2^2(\phi)})^{4/3}.
\end{eqnarray}
Hence we can focus on the integral
\begin{equation}\label{null2}
\int_{\sigma}T^{\text{eff}}_{\mu\nu}k^\mu k^\nu dl=\text{C}\int^{+\infty}_{\phi_0}(\frac{1-\gamma^2\Omega_2^2(\phi)} {4(1+\gamma^2)(1+\gamma^2\Omega_2^2(\phi))}-\frac{1-3\gamma^2\Omega_2^2(\phi)}{2 (1+\gamma^2\Omega_2^2(\phi))^2})(\frac{|\Omega_1(\phi)|}{1+\gamma^2\Omega_2^2(\phi)})^{4/3}d\phi,
\end{equation}
with $\text{C}:=-\frac{9\text{sgn}(p_\phi)}{32\pi^2G\gamma^2\rho_{\text{c}}^{\text{TR}}\Delta^2 \sqrt[3]{|p_\phi|}}(\sqrt{\frac{4\pi G\Delta}{3}})^{2/3}$ being a constant.
%\begin{equation}\label{null3}
%\int_{\gamma}T^{\text{eff}}_{\mu\nu}k^\mu k^\nu dl=\text{C}\int_{-\infty}^{\phi_0}(\frac{1-\gamma^2\Omega_2^2(\phi)} {4(1+\gamma^2)(1+\gamma^2\Omega_2^2(\phi))}-\frac{1-3\gamma^2\Omega_2^2(\phi)}{2 (1+\gamma^2\Omega_2^2(\phi))^2})(\frac{|\Omega_1(\phi)|}{1+\gamma^2\Omega_2^2(\phi)})^{4/3}d\phi.
%\end{equation}
Let us take $\phi_0=0$ without loss of generality, the above integrals is given by
\begin{equation}\label{nullresult1}
\int_{\sigma}T^{\text{eff}}_{\mu\nu}k^\mu k^\nu dl=0.0584303\text{C}<0
\end{equation}
 for $\gamma=0.273985635$, or given by
 \begin{equation}\label{nullresult2}
\int_{\sigma}T^{\text{eff}}_{\mu\nu}k^\mu k^\nu dl=0.0674933\text{C}<0
\end{equation}
 for $\gamma=0.23753295796592$.
\subsection{The dominant and strong energy condition}
Recall the FRW metric $ds^2=-dt^2+a^2(dx^2+dy^2+dz^2)$, we can choose an orthogonal and normalized basis which is given by
\begin{equation}
dt'=dt,\quad dx'=adx,\quad dy'=ady,\quad dz'=adz.
\end{equation}
Then, the effective energy and momentum tensor takes the formulation $T^{\text{eff}}_{\mu\nu}=\rho_{\text{eff}}(dt')_\mu(dt')_\nu+P_{\text{eff}}((dx')_\mu(dx')_\nu+(dy')_\mu(dy')_\nu+(dz')_\mu(dz')_\nu)$. The dominant and strong energy conditions require
\begin{equation}
\rho_{\text{eff}}\geq|P_{\text{eff}}|\quad(dominant \quad energy \quad condition)
\end{equation}
and
\begin{equation}\label{strong}
\rho_{\text{eff}}+P_{\text{eff}}\geq0\ \text{and}\ \ \rho_{\text{eff}}+3P_{\text{eff}}\geq0, \quad(strong\quad energy \quad condition)
\end{equation}
respectively. In order to simplify our calculation, let us reformulate the effective energy density and pressure by using Eqs.\eqref{111}, \eqref{444} and \eqref{555}, which read
\begin{eqnarray}
P_{\text{eff}}
%&=&-\frac{3}{8\pi\gamma^2\Delta}(\frac{\rho_\phi(1-\gamma^2\Omega_2^2(\phi))} {2\rho_{\text{c}}^{\text{TR}}(1+\gamma^2)(1+\gamma^2\Omega_2^2(\phi))}-\frac{\rho_\phi(1-\gamma^2\Omega_2^2(\phi))}{\rho_{\text{c}}^{\text{TR}} (1+\gamma^2\Omega_2^2(\phi))^2}+\frac{\gamma^2\Omega_2^2(\phi)}{ (1+\gamma^2\Omega_2^2(\phi))^2}(1+\frac{\rho_\phi}{ \rho_{\text{c}}^{\text{TR}}}))\\\nonumber
&=&-\frac{3}{8\pi\gamma^2\Delta}(\frac{\rho_\phi(1-\gamma^2\Omega_2^2(\phi))} {2\rho_{\text{c}}^{\text{TR}}(1+\gamma^2)(1+\gamma^2\Omega_2^2(\phi))}-\frac{\rho_\phi(1-2\gamma^2\Omega_2^2(\phi))}{\rho_{\text{c}}^{\text{TR}} (1+\gamma^2\Omega_2^2(\phi))^2}+\frac{\gamma^2\Omega_2^2(\phi)}{ (1+\gamma^2\Omega_2^2(\phi))^2})
\end{eqnarray}
and
\begin{equation}
\rho_{\text{eff}}=\frac{3}{8\pi\gamma^2\Delta}(\frac{\gamma^2\Omega_2^2(\phi)}{ (1+\gamma^2\Omega_2^2(\phi))^2}(1-\frac{\rho_\phi}{\rho_{\text{c}}^{\text{TR}}})).
\end{equation}

Now let us discuss the dominant and strong energy condition respectively.
First, notice that the dominant energy condition $\rho_{\text{eff}}\geq|P_{\text{eff}}|$ equivalent to $\rho_{\text{eff}}\geq P_{\text{eff}}\geq -\rho_{\text{eff}}$,
which can be expressed explicitly as
\begin{eqnarray}\label{eq1}
P_{\text{eff}}\geq -\rho_{\text{eff}}\   \mapsto \quad -\gamma^4\Omega_2^4(\phi)+6(1+\gamma^2)\gamma^2\Omega_2^2(\phi)-2\gamma^2-1\leq0,\
\end{eqnarray}
and
\begin{eqnarray}  \label{eq2}
 \rho_{\text{eff}}\geq P_{\text{eff}}\   \mapsto\quad \frac{\rho_\phi(-\gamma^4\Omega_2^4(\phi)+2(1+\gamma^2)\gamma^2\Omega_2^2(\phi)-2\gamma^2-1)} {2\rho_{\text{c}}^{\text{TR}}(1+\gamma^2)}+2\gamma^2\Omega_2^2(\phi)\geq0.
\end{eqnarray}
 These two equations can be discussed separately.
 The first equation \eqref{eq1} requires that $3(1+\gamma^2)-\sqrt{9(1+\gamma^2)^2-(2\gamma^2+1)}\geq\gamma^2\Omega_2^2(\phi)$ or $\gamma^2\Omega_2^2(\phi)\geq3(1+\gamma^2)+\sqrt{9(1+\gamma^2)^2-(2\gamma^2+1)}$.  % Notice that the minimal value of $\Omega_2^2(\phi)$ tends to $1$ and we have $3(1+\gamma^2)-\sqrt{9(1+\gamma^2)^2-(2\gamma^2+1)}>\gamma^2$.
 By using $\gamma=0.23753295796592$, we have $\Omega_2^2(\phi)\leq3.20302326$ or $\Omega_2^2(\phi)\geq109.138644$. Or by using $\gamma=0.273985635$, we have $\Omega_2^2(\phi)\leq2.44479315$ or $\Omega_2^2(\phi)\geq83.4826008$. Also, it can be verified that the second equation \eqref{eq2} always holds when \eqref{eq1} is satisfied (see Figure 1 and 2). Now, we conclude that the dominant energy condition is satisfied if $\Omega_2^2(\phi)\leq3.20302326$ or $\Omega_2^2(\phi)\geq109.138644$ for $\gamma=0.23753295796592$, while if $\Omega_2^2(\phi)\leq2.44479315$ or $\Omega_2^2(\phi)\geq83.4826008$ for $\gamma=0.273985635$.

Similar calculation can be given for the strong energy condition \eqref{strong}. The first equation in Eqs.\eqref{strong} requires
\begin{eqnarray}\label{strongeq1}
\rho_{\text{eff}}+P_{\text{eff}}&=&-\frac{3\rho_\phi}{8\pi\gamma^2\Delta\rho_{\text{c}}^{\text{TR}}}(\frac{(1-\gamma^2\Omega_2^2(\phi))} {2(1+\gamma^2)(1+\gamma^2\Omega_2^2(\phi))}-\frac{(1-3\gamma^2\Omega_2^2(\phi))}{ (1+\gamma^2\Omega_2^2(\phi))^2})\\\nonumber
&=&\frac{3\rho_\phi}{8\pi\gamma^2\Delta\rho_{\text{c}}^{\text{TR}}}\frac{\gamma^4\Omega_2^4(\phi) -6(1+\gamma^2)\gamma^2\Omega_2^2(\phi)+(1+2\gamma^2)} {2(1+\gamma^2)(1+\gamma^2\Omega_2^2(\phi))^2}\\\nonumber
&\geq&0,
\end{eqnarray}
which holds if $3(1+\gamma^2)-\sqrt{9(1+\gamma^2)^2-(2\gamma^2+1)}\geq\gamma^2\Omega_2^2(\phi)$ or $\gamma^2\Omega_2^2(\phi)\geq3(1+\gamma^2)+\sqrt{9(1+\gamma^2)^2-(2\gamma^2+1)}$. By using $\gamma=0.23753295796592$, we have the corresponding numerical solution   $\Omega_2^2(\phi)\leq3.20302326$ or $\Omega_2^2(\phi)\geq109.138644$. While by using $\gamma=0.273985635$, we have $\Omega_2^2(\phi)\leq2.44479315$ or $\Omega_2^2(\phi)\geq83.4826008$. The second equation in Eqs.\eqref{strong} requires
\begin{eqnarray}
&&\rho_{\text{eff}}+3P_{\text{eff}}\\\nonumber
&=&-\frac{3}{8\pi\gamma^2\Delta}(\frac{3\rho_\phi(1-\gamma^2\Omega_2^2(\phi))} {2\rho_{\text{c}}^{\text{TR}}(1+\gamma^2)(1+\gamma^2\Omega_2^2(\phi))}-\frac{\rho_\phi(3-7\gamma^2\Omega_2^2(\phi))}{\rho_{\text{c}}^{\text{TR}} (1+\gamma^2\Omega_2^2(\phi))^2}+\frac{2\gamma^2\Omega_2^2(\phi)}{ (1+\gamma^2\Omega_2^2(\phi))^2})\\\nonumber
%&=&-\frac{3}{8\pi\gamma^2\Delta}(\frac{3\rho_\phi(1-\gamma^4\Omega_2^4(\phi)-2(1+\gamma^2)(1-7/3\gamma^2\Omega_2^2(\phi)))+4(1+\gamma^2)\rho_{\text{c}}^{\text{TR}} \gamma^2\Omega_2^2(\phi)} {2\rho_{\text{c}}^{\text{TR}}(1+\gamma^2)(1+\gamma^2\Omega_2^2(\phi))^2})\\\nonumber
%&=&-\frac{3}{8\pi\gamma^2\Delta}(\frac{\frac{9}{8\pi G\Delta}(\Omega_2^2(\phi)-1)\left(1-\gamma^4\Omega_2^4(\phi) -2(1+\gamma^2)(1-7/3\gamma^2\Omega_2^2(\phi))\right)+4(1+\gamma^2)\rho_{\text{c}}^{\text{TR}} \gamma^2\Omega_2^2(\phi)(1+\gamma^2\Omega_2^2(\phi))^2} {2\rho_{\text{c}}^{\text{TR}}(1+\gamma^2)(1+\gamma^2\Omega_2^2(\phi))^4})\\\nonumber
&\geq&0,
\end{eqnarray}
which can be solved for $\Omega_2^2(\phi)$ by numerical method.
By using $\gamma=0.23753295796592$, we have that the above equation holds if $\Omega_2^2(\phi)\geq146.921150827$. By using $\gamma=0.273985635$, we have that the above equation holds if  $\Omega_2^2(\phi)\geq112.2645175$. Then, notice the solution of Eq.\eqref{strongeq1} we can conclude that the strong energy condition holds if $\Omega_2^2(\phi)\geq 146.921150827$ for $\gamma=0.23753295796592$, or if $\Omega_2^2(\phi)\geq 112.2645175$ for $\gamma=0.273985635$.

Now it is ready to have an overall look on the violation of dominant and strong energy condition. Recall that the bounce happens at the point $\Omega_2^2(\phi)= \frac{1+2\gamma^2}{\gamma^2}$. It is easy to see that the dominant energy condition with respect to the effective stress-energy tensor in the new model of LQC  is violated at the period $3.20302326\leq\Omega_2^2(\phi)\leq109.138644$ (for $\gamma=0.23753295796592$) or $2.44479315\leq\Omega_2^2(\phi)\leq83.4826008$ (for $\gamma=0.273985635$) around the bounce point, while the strong energy condition  is violated at the period $\Omega_2^2(\phi)\leq 146.921150827$ (for $\gamma=0.23753295796592$) or  $\Omega_2^2(\phi)\leq 112.2645175$ (for $\gamma=0.273985635$).
 Hence, we can conclude that the strong energy condition is violated not only at a period around the bounce point, but also the whole period from the bounce point to the classical phase corresponding to the de Sitter epoch.
All of these results for $\gamma=0.23753295796592$ are illustrated in the Figure 1 and Figure 2.
\begin{figure}[h]
 \includegraphics[scale=0.59]{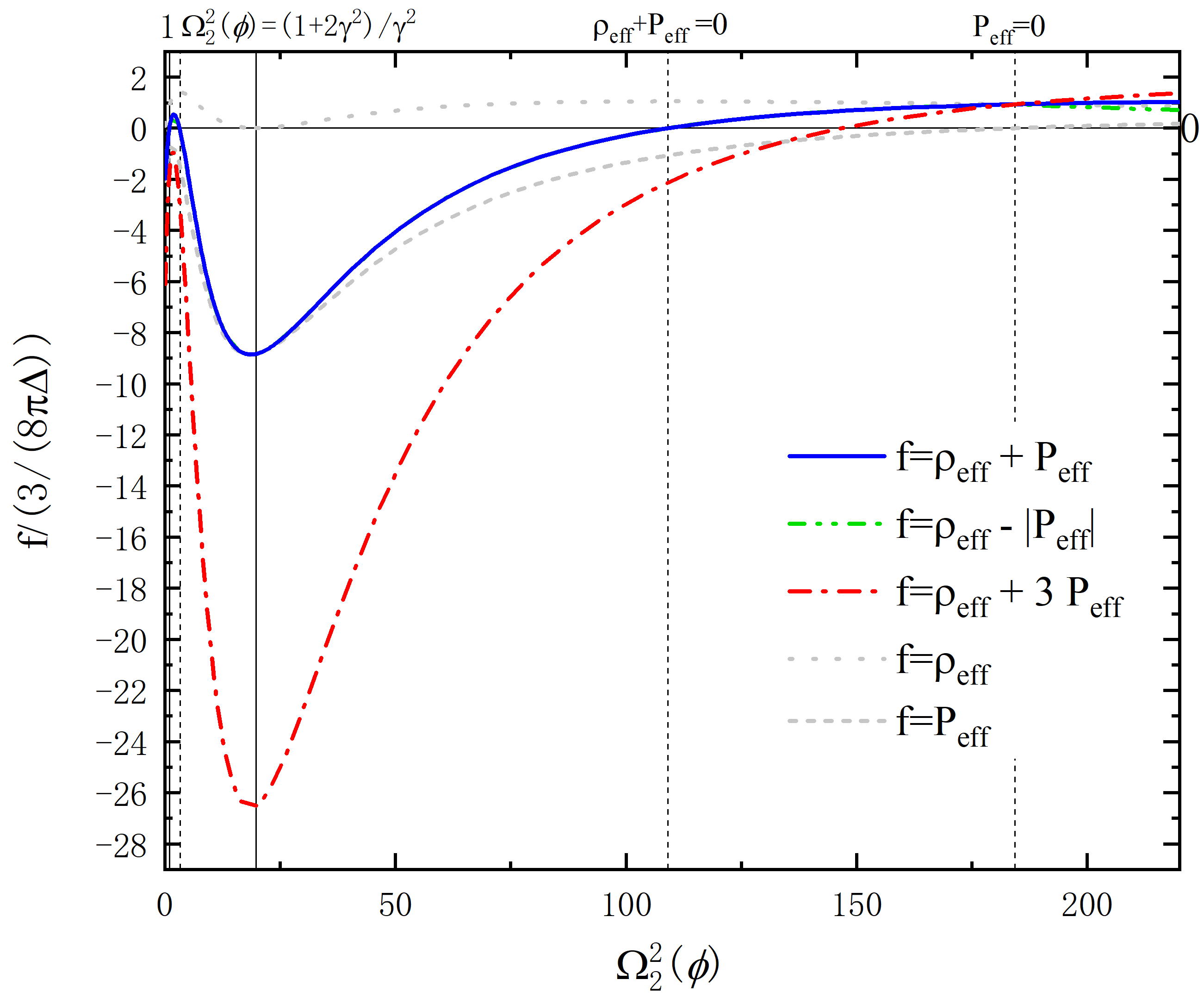}
\caption{The behaviour of $\rho_{\text{eff}}$, $P_{\text{eff}}$ and energy conditions for $\gamma=0.23753295796592$ and $1\leq\Omega_2^2(\phi)\leq250$.}
\label{fig:label1}
\end{figure}
\begin{figure}
 \includegraphics[scale=0.59]{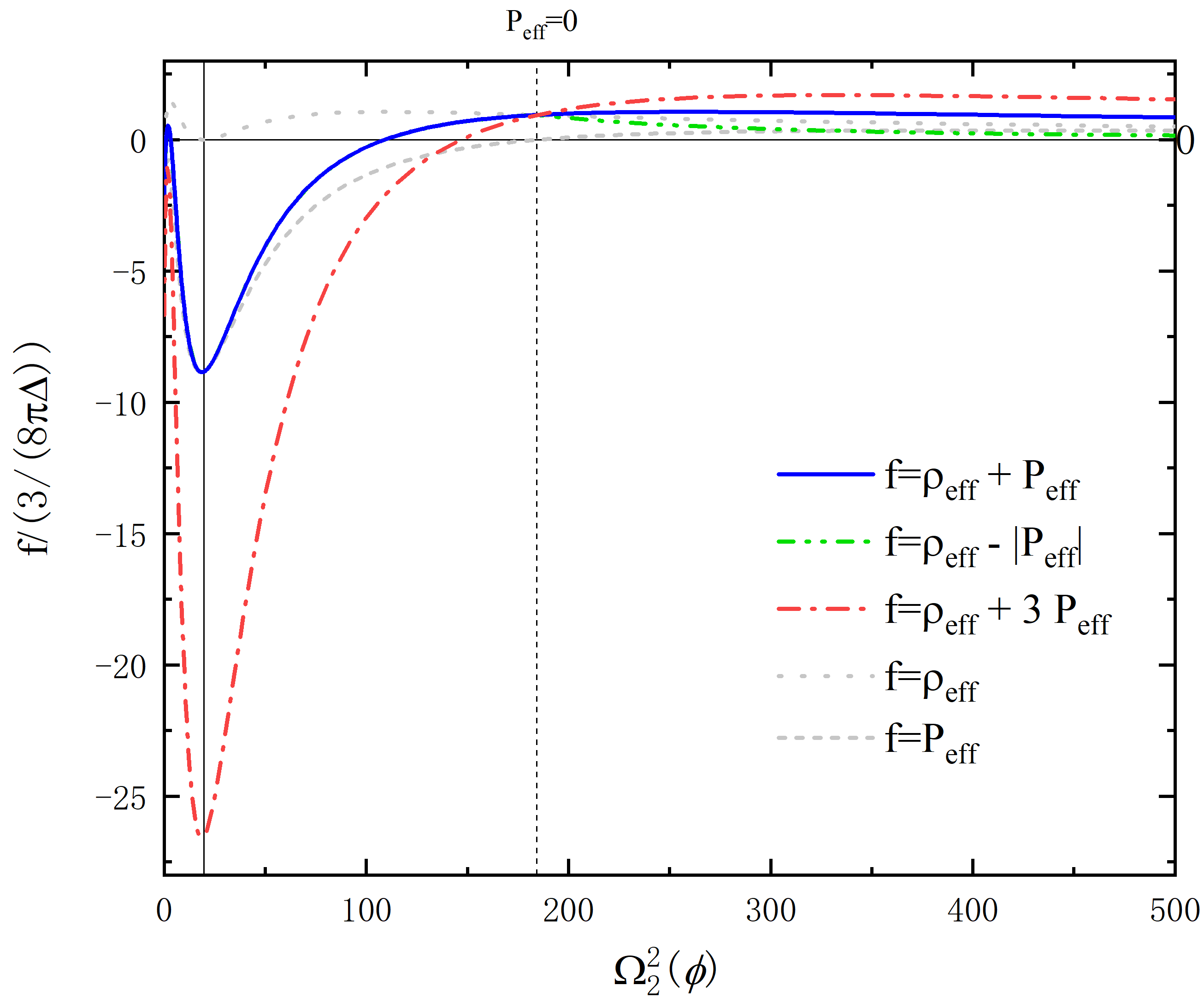}
\caption{The behaviour of $\rho_{\text{eff}}$, $P_{\text{eff}}$ and energy conditions for $\gamma=0.23753295796592$ and $1\leq\Omega_2^2(\phi)\leq500$.}
\label{fig:label2}
\end{figure}
One can see that there are the two branches of evolution of universe divided by the bounce point at $\Omega_2^2(\phi)= \frac{1+2\gamma^2}{\gamma^2}=6.2099421$ for $\gamma=0.23753295796592$. The effective pressure $P_{\text{eff}}$ is always negative in the branch which contains the de Sitter epoch, while it is negative only in a period near the bounce point in another branch.
\section{Conclusion and discussion}
The new model of loop quantum cosmology gives a remarkable prediction of the emergent of de Sitter epoch, in which the large cosmology constant from quantum geometry correction is able to influence the behaviour of the energy conditions obviously. We studied the interesting violation of energy condition problem in this model and considered the average null, dominant and strong energy conditions with respect to the effective stress-energy tensor given by the effective dynamics equations. The results shown that the quantum correction has a significant influence on the behaviours of violation of energy condition.

Based on the effective Hamiltonian constraint in the new model of LQC, we gave the effective dynamics equations and corresponding effective stress-energy tensor by comparing with the standard Friedmann equation and Raychaudhuri equation. Then, the averaging null, dominant and strong energy conditions with respect to the effective stress-energy tensor were expressed as functions of the physical time $\phi$. The numerical calculation showed that, with respect to the effective stress-energy tensor in the new model of LQC, the averaging null energy condition is violated while the dominant energy condition is violated only at a period around the bounce point. Last but not least, the strong energy condition is violated not only at a period around the bounce point, but also the whole period from the bounce point to the classical phase corresponding to the de Sitter period. Such a result is consistency with the appearing of the de Sitter epoch in the new model of LQC, in which the large effective cosmology constant is inversely proportional to the smallest non-zero area in LQG.

In fact, the violation of the averaging null and strong energy conditions with respect to the effective stress-energy tensor has been approved in the standard LQC \cite{2009Averaged}\cite{Xiong:2006ey}, in which the strong energy conditions is only violated near the bounce point comparing the new model of LQC.
 The new model of LQC introduces the de-Sitter epoch with a large effective cosmology constant coming from the loop quantum effects, which contributes to the violation of energy conditions additionally. This result indicates us that the cosmology constant in our true universe may have a quantum gravity origin. However, the astronomical observation shows that the cosmology constant in our true universe is small, hence it is fail to use the de-Sitter epoch in the new model of LQC to describe the far further of the true universe as well as to find a period to describe the present true universe in the new model of LQC. Generally, though the new model of LQC do not predict the correct cosmology constant, it still provides us a new perspective to describe the origin of the cosmology constant, and it is expected to extend the core idea of this new model to other loop quantum gravity theories, i.e., loop quantum $f(R)$ theory and higher dimensional LQG \cite{2011Loop,Zhang:2011vi,Bodendorfer:Qu,Long:2019nkf,Zhang:2015bxa},  to find a proper quantum gravity theory which could predict the accurate cosmology constant in further research. Moreover, by sharing the same quantum geometry nature, our results is expected to be inherited in the full loop quantum gravity so that some lights are shed on construction of the wormhole and time machine which usually need exotic matters violate energy conditions.
\section*{Acknowledgments}
This work is supported by the National Natural Science Foundation of China (NSFC) with Grants No. 11775082, No. 11875006 and No. 11961131013.
\bibliographystyle{unsrt}

\bibliography{ref}

%\appendix
%\section{Graphics illustration of the behaviours of energy conditions with respect to the effective stress-energy tensor in new model of LQC}

%\begin{figure}[h]
 %\includegraphics[scale=0.59]{fig2St_gm027_0_180.png}
%\caption{The behaviour of $\rho_{\text{eff}}$, $P_{\text{eff}}$ and energy conditions for $\gamma=0.273985635$ and $1\leq\Omega_2^2(\phi)\leq180$.}
%\label{fig:label3}
%\end{figure}

%\begin{figure}[h]
 %\includegraphics[scale=0.59]{fig2St_gm027_0_500.png}
%\caption{The behaviour of $\rho_{\text{eff}}$, $P_{\text{eff}}$ and energy conditions for $\gamma=0.273985635$ and $1\leq\Omega_2^2(\phi)\leq500$.}
%\label{fig:label4}
%\end{figure}

\end{document}